# Trust and distrust in electoral technologies: what can we learn from the failure of electronic voting in the Netherlands (2006/07)

DAVID, DUENAS-CID

Management in Networked and Digital Societies, Kozminski University, Warsaw, Poland, dduenas@kozminski.edu.pl

**Abstract.**

This paper focuses on the complex dynamics of trust and distrust in digital government technologies by approaching the cancellation of machine voting in the Netherlands (2006-07). This case describes how a previously trusted system can collapse, how paradoxical the relationship between trust and distrust is, and how it interacts with adopting and managing electoral technologies. The analysis stresses how, although being a central component, technology's trustworthiness dialogues with the socio-technical context in which it is inserted, for example, underscoring the relevance of public administration in securing technological environments. Beyond these insights, the research offers broader reflections on trust and distrust in data-driven technologies, advocating for differentiated strategies for building trust versus managing distrust. Overall, this paper contributes to understanding trust dynamics in digital government technologies, with implications for policymaking and technology adoption strategies.

### CCS CONCEPTS

Trust • Voting / election technologies • Management of computing and information systems

### Keywords

Trust, Distrust, Machine voting, Netherlands

## 1 Introduction

Numerous researchers highlighted the central role of trust in technology adoption [4] [5] [69], particularly within the field of digital government [9]. The intuitive connection between trusting citizens and their propensity to embrace government-led innovations [11], coupled with the ease of adoption for technologies perceived as trustworthy [8], seems evident. Factors such as transparency, openness, and the ability to involve citizens in co-creating technological advancements are acknowledged contributors to building trust and, consequently, positively impact citizen engagement.

Yet, what happens when established technologies face a crisis of trust and orchestrated campaigns sow seeds of distrust? This paper, similarly to Olesk's approach [65], delves into the flip side of the equation, examining a well-known case from the electronic voting literature—the machine voting abandonment in the Netherlands (2006-07) [31] [45] [53]. By revisiting this case, the research approaches the intricate dynamics of trust erosion and the paradoxical relationship between trust and distrust. This exploration sheds light on how distrust is created and offers valuable insights applicable to other digital government technologies.

## 2 Theoretical framework

Trust is a complex concept encompassing many nuances and interpretations [21]. While being a widely used concept in everyday conversations, it also attracted the attention of various academic disciplines, particularly within the social sciences. Disciplines such as sociology [25] [48] [63], psychology [2] [22] [64], law [13] [29] [66], or management [30] [35] [62] have extensively explored the meanings and understanding of trust. Moreover, trust has emerged as a central concept in the field of information studies and e-government research, playing a crucial role in technology adoption theories and models such as TAM or UTAUT.

In contrast, the concept of distrust has not attracted the same level of attention among scholars [59], often being viewed simply as the symmetric opposite of trust. Following the insights of Luhmann [48] and Lewicki [41], I understand distrust as a concept possessing its own distinct patterns of functioning, closely intertwined with but separate from trust: not trusting does not necessarily mean distrusting. Trust can be perceived as a passive element rooted in the absence of evidence of untrustworthiness, whereas distrust is actively constructed through concerted efforts. This perspective aligns with the findings of Kühne [39], who observed that trust relies on the absence of active demonstrations of untrustworthiness. Conversely, distrust tends to perpetuate itself and can extend its influence to adjacent targets, as noted by D'Cruz [15], highlighting its propensity for self-perpetuation and spill-over effects [33]. Recognizing the nuanced distinction between trust and distrust allows understanding that individuals may simultaneously place trust in certain aspects of a system while distrusting others (e.g., distrusting politicians while trusting democracy [63]), that trust and distrust can fluctuate over time, even towards the same entity, or that certain elements within a system may foster both trust and distrust at the same time.

However, eGovernment research has largely overlooked this differentiation. Much of the scholarly work has approached trust as an isolated concept, often viewed solely as a prerequisite for technology adoption [11] [44] or as a consequence of implementing data-driven technologies within institutions [32] [57]. Of particular concern is the manner in which trust is integrated into models such as the Technology Acceptance Model (TAM) [34] [67] and the Unified Theory of Acceptance and Use of Technology (UTAUT) [3] [12]. In these frameworks, trust is often positioned as either a precursor to adoption or as an intermediary determinant of behavioral intention. This oversimplification has been criticized by Shachak [58] since the use of TAM and UTAUT models has reached a situation in which some concepts are used in a simplistic manner, not allowing achieving a full comprehension of trust's potential impact. Similarly, Williams et al. [68] raise concerns about the utilization of UTAUT, arguing that while the model's prominence in Information Systems research is indisputable, its conceptualization of key elements, such as trust, remains underdeveloped. When using those models, trust is often mentioned but not thoroughly explored, thereby limiting the advancement of knowledge in this domain.

This paper advocates for treating trust and distrust as distinct yet interconnected concepts in light of these limitations. When applied to digital contexts, it is necessary to consider the role of technology in reshaping our understanding of trust and distrust, as well as their respective formation processes [21]. By embracing this perspective, I aim to contribute to a more nuanced and comprehensive understanding of trust and distrust dynamics in digital governance. For doing so, this paper focuses on a specific domain within eGovernment: the utilization of technology in electoral management. Although election-related topics are often approached from political sciences, the topic is not new in technology and information studies [6] [24] [56], especially since technologies have become an important element for electoral management [1] [43]. However, elections possess distinct characteristics that make it particularly relevant for exploring trust and distrust dynamics [19]:

- Time constraints: Elections operate on a cyclical schedule rather than providing an ongoing service. Each election is organized from scratch, relying heavily on past experiences and established organizational frameworks. Innovations and technological implementations should be ready for working on a specific date, as repeating the process incurs substantial costs.
- Structural complexity: Given the non-permanent nature of electoral management, most participants and stakeholders involved in its organization often work partially on this issue, but they invest a large number of hours in a short period of time [36] creating a temporary intensive network of relations [51]. While crucial for effective electoral management, this large number of temporary actors may not always possess a comprehensive understanding of the intricacies and demands of their roles and, especially, of the roles of others.
- Integration Challenges: The integration of technology into electoral processes amplifies management complexity. It entails not only adding a technological layer but integrating it into existing structures to deliver a unified outcome within tight timeframes [38]. Harmonizing diverse stakeholders has been a recurring problem in successful adoption of electoral technology [18].
- Political significance: Elections represent the cornerstone of democracy, crucial for delegating power. Consequently, they are subject to intense political scrutiny to ensure an accurate transmission of results and facilitate the acceptance of the losers' results [20]. Trust, hence, encompasses a broader political frame than election management.

The preceding elements help us understand not only the particularities of electoral management within the context of public administration but also the significance of trust- and distrust-related dynamics when introducing technologies into electoral processes. Electoral technologies are distributed throughout the electoral cycle [37] encompassing both visible and invisible components. While certain technologies, such as digital voter lists or result transmission systems, operate behind the scenes for managerial or administrative purposes, others intersect directly with citizens, particularly during the process of identification or vote casting. Although this paper will focus on the latter, I acknowledge the relevance of researching trust- and distrust-related elements regarding digital identification systems, widely implemented in some African countries under the promise of improving the integrity of elections [16] [23] [52].

Debates surrounding the trustworthiness of voting methods have particularly intensified in relation to vote casting. Various forms of electronic and internet voting have been implemented worldwide, with some countries such as Estonia [60], Switzerland [50], Brazil [42], Belgium [14] or India [17] successfully adopting these methods. However, these systems often face scrutiny regarding their security and trustworthiness [28] [61]. In this paper, the focus will be directed towards the discontinued use of electronic voting in the Netherlands after 2007.

Voting machines were introduced in the Netherlands in 1966, initially adopted by two municipalities [46], and gradually expanded in usage until 2006, when a large majority of Dutch citizens were casting their votes using electronic voting machines [45]. Their adoption aimed to streamline the cumbersome process of manual vote counting, enhance accuracy, and mitigate human errors in vote casting, which often resulted in invalidated ballots [10]. However, the credibility of voting machines was challenged by the hacktivist group "Wij vertrouwen stemcomputers niet" (Trad. We don't trust voting computers). The group demonstrated that the voting machines were susceptible to hacking and compromising the secrecy of the vote [26]. Their successful campaign exposed a set of vulnerabilities provoking the decision to discontinue the use of voting machines in 2008, reverting to paper ballots—a decision that remains unchanged to date.

## 3   Method

Using the Dutch case as a frame, the paper approaches the dynamics of trust and distrust creation in electoral technologies. The reasons for choosing this case are the following: 1) exists a robust body of academic literature providing a solid foundation for examining the case, 2) despite the existing literature, the proposed angle of this paper has not been thoroughly explored, 3) key actors involved in the case remain accessible and willing to discuss the issue once emotions surrounding the event are over, and 4) the type of research question proposed invites for a retrospective analysis.

As previously suggested, the debate on trust and distrust in technology is often biased towards the technological and technical questions, overlooking critical dimensions of human behavior such as interactions between humans and/or organizations. Most of the requirements that electoral technologies must meet are related to the creation of a trustworthy system, encompassing security, privacy protection, and verifiability [55]. At the same time, the perception of trust must permeate the voter community, with various stakeholders contributing to shaping this perception of security and trustworthiness.

By examining the Dutch Voting Machine crisis of 2006-07, this paper proposes a qualitative remapping of the scenario, aiming to gather and analyze the discourses of key actors involved in the conflict. Through this approach, the paper seeks to identify the primary elements influencing the creation of trust and distrust and their potential impact on public perception.

The research methodology involves qualitative interviews conducted between August 16 and September 6, 2021, during a study visit to The Netherlands hosted by Vrije University. The interviews engaged members of the hacktivist group, the electoral management board, the Ministry of Interior Affairs, researchers in IT and social sciences, electoral observers, and journalists—all directly involved in the crisis. Notably, the producers of the technology were impossible to contact for interviews. While the number of interviews conducted may seem modest for a qualitative study, the significance of the interviewees and their direct involvement in the events analyzed mitigates this potential limitation. The qualitative research was supplemented by a comprehensive review of available literature on the case. Coding and analysis of the 525 minutes of interview recordings was conducted using NVivo software.

## 4   Results

This section describes a significant sequence of the facts that occurred, presenting different arguments given for trusting or distrusting voting machines. It unveils a paradoxical scenario wherein trust was posited in a not trustworthy system, and even when its lack of trustworthiness was exposed, it failed to significantly alter public perception regarding its suitability. This situation draws a landscape in which arguments for trusting or distrusting overlap each other, having differential impacts on the public perception of the system. Based on the interviews collected, a set of arguments will be discussed, divided into three moments (before, during and after the voting machine crisis) and including theoretical references to previous research on the topic.

### 4.1   The Dutch Context.

Regarding election management, the features of the Dutch political landscape contribute to keeping high levels of trust in the election outcomes by reducing the political tension, *"I think that's partly also because of our multi-party system where everybody who is a serious contender will end up with a with a seat in Parliament. So, everybody feels represented, nobody feels robbed of the election"* (Int. 8). The proportional representation system used in Dutch elections, with a variable threshold calculated by dividing the number of votes cast by the seats of the Parliament (generally between 0,65-0,7%), makes it easy for new parties to access the Parliament. In contrast, this avoids having close competition between a reduced number of parties and, therefore, plays a role in avoiding negative perceptions towards the election outcomes, *"the term election fraud in the Dutch history has never been a hot topic"* (Int. 8). In this context, where the number of critical voices regarding the electoral process is reduced to a small number of highly politically motivated citizens, the emergence of the hacktivist group stood out as a surprise. (*"We were surprised not understanding what (hacktivist name) wanted with it… (…) who are these troublemakers? And not understanding that they were really worried (…) that something bad could happen with the election. So that they were doing it out of concern for democracy"*, Int. 6)

### 4.2   Adopting the technology.

Practical considerations primarily drove the motivations behind the adoption of voting machines; they were "*very pragmatic*" (Int.7), searching for convenience and efficiency in the electoral process. In the Netherlands, elections are managed by local administrations, with municipalities being the primary beneficiaries of the use of voting machines. Their adoption reduced the task of ballot counting, significantly streamlining the electoral process and enabling more efficient resource allocation (*"For municipalities it was great because it made the organization of the election so easy,"* Int.6), and reducing the possible human mistakes in the counting process *"we don't need to have all these people doing the counting (…) they make mistakes with the counting… you (could) hear this many, many times…"* (Int. 4). In a general election, a ballot can contain over 600 candidates and just one receives a vote, by ticking with a pen in the correct box in the ballot. Votes are counted by hand.

Differently, for citizens the increase in convenience was not so evident. While voting machines seemed to facilitate the act of voting itself (*"the general public was surprised that the voting computers disappeared and found voting on paper more cumbersome,"* Int.2), other aspects typically associated with voting convenience, such as the need to travel long distances to cast a ballot, were considered irrelevant (*"The Netherlands are not a big country. It is a small country (…) it has good railroads, roads, all the Dutchman know how to ride a bike. There are 9000 polling stations on Election Day, every Dutchman who rides a bike, within five minutes, can reach a polling station"* Int.3).

Moreover, the rapid transmission of results facilitated the extension of polling station hours until late in the day (the Netherlands conducts elections on weekdays) *"to give people who work the opportunity to go and vote"* (Int. 7), and provide the electoral results on an early hour, and *"the elected politicians thought it was great because the result of the election was immediately after closing the ballot papers"* (Int. 6).

### 4.3 The blinding relation between trust and convenience.

Trust has been recognized as a mechanism for reducing the complexity of social interaction [48]. As such, while the convenience offered by technology stands out as an argument for its adoption [40], it also contributes to masking potential sources of distrust under the assumption that the technology operates smoothly. While in other forms of electronic voting, such as internet voting, convenience primarily benefits users by reducing the transaction costs associated with voting [27], in the case of Dutch voting machines, convenience is closely tied to the management of elections. The use of voting machines mainly eased the counting process *("there are about 26 parties competing in the elections (...) our ballot paper (...) it's like this (big) (...) you first have to unfold all this huge paper and then have to find where is the vote in the ballot..."*, Int.3) and the general management of elections (*"A lot of the people on the municipal level running elections still are (...) pushing for going back. They were angry because it's a lot more work"*, Int. 8).

Furthermore, the cost-effectiveness of the system contributed to its widespread adoption, "it wasn't very costly, because the voting machine… some municipalities used them for ten years or longer, because they were quite robust" (Int. 5), offering an attractive solution for reducing the organizational burden of electoral administration. But, as suggested previously, the system's convenience, together with a record of lack of problems, can be factors that help reduce the scrutiny of the system. "In the beginning, people are critical and ask questions, and then after ten years, there's this sense of complacency and we don't know really what's going on anymore, but that's fine because we've used it for ten years and went well, so we can trust it and everybody kind of nods off and move on to other questions… and we lose the scrutiny that is, I think, fundamental when you are holding elections because it's such an integral part of the workings of e-democracy" (Int. 8). Overall, the prevailing perception among the public was that "voting computers, even the ones that were in use in the Netherlands, where really OK because they worked, and they were easy to use and all that" (Int. 2)

### 4.4 Not everyone agrees.

The adoption of voting machines in the city of Amsterdam, one of the last to introduce them, generated a reaction against their use by a group of hacktivists, *"During the municipal elections of March 2006, Amsterdam was, for the first time, voting by voting computer (...) and this led to the campaign by (hacktivist), who was in for the first time confronted in the polling station where the computer and said I don't trust this"* (Int. 8). The hacktivist leader was well known in the country for its previous activities and got some legitimacy within the expert community *"because he had also studied the phenomenon of digital voting, and he decided that there was something very wrong with it. And because (name) had some notoriety, he was a well-known hacker. He had done things before (...) if (name) is worried about something, I should take it seriously"* (Int. 2).

The main arguments against voting machines were concentrated around two ideas: the incapacity of machines to combine transparency and secrecy *("We need transparency, and we need secrecy of the votes (...) we cannot do away with either of these two (...) Given that we can't do away with either of these properties, we are forced to have physical artifacts representing each vote. And we have to be able to manually count, at least, some significant fraction of these votes in order to have public trust in elections",* Int. 5) and the black-box ownership system *("it's not a great idea that we've outsourced our elections to a company and done it in a black box (...) you enter something in the computer and then, at the end of the day, you press result and you don't know if what's been entered during the day is actually what comes out"* Int. 8). Technically, that translated into the possibility of hacking the computer software (although two companies were providing the software, both Dutch-owned, most of the municipalities were using NEDAP computers, contributing to a one-company dependency) to favor certain parties without being noticed, and the possible secrecy breach by the transmission of identifiable frequencies [26].

### 4.5 Campaign against voting machines.

The vulnerabilities of the voting machines came to light when the hacktivist group "We Don't Trust Voting Computers" revealed their lack of trustworthiness due to technical deficiencies. It helped that the voting computers *"were essentially designed in the late 80s, early 90s"* (Int. 5), when *"the whole issue of computer security was not on the agenda yet"* (Int. 7). The technical weaknesses of the voting machines were surrounded by an inadequate organizational framework. The government had outsourced the production of these machines, with no oversight on the software used *("the government didn't even see the software,"* Int. 5). Instead, the government relied on a certification company to ensure that the machines complied with legal requirements, but "*the law said nothing about computer security… about security against manipulation"* (Int. 5). The regulations at the time *"were focused on robustness, so* (for example) *there was a requirement that you could drop* (water) *down and still work"* (Int. 6). In addition, the lack of regulation on how the computers should be stored allowed the hacktivists to access them and manipulate the software. While the regulation included detailed provisions for the management of paper ballots, almost nothing was said concerning voting machines.

This lack of stringent regulations left the control of the system's inner workings mainly in the hands of the voting computers provider, *"the thread model* (was) *outsiders should not be able to defraud the election and insiders have free rein"* (Int. 5). In other words, if the computer provider attempted to manipulate the results, there were few safeguards in place to prevent it, *"Nobody really knew what went on after you push the button (...) the manufacturer of the voting machines didn't want to share that with anyone else, because it has commercial value to not to disclose how this works"* (Int. 3).

The hacktivist group conducted a successful campaign unveiling the weaknesses of the system, fueled by their participation in a TV show where they show how results could be manipulated and proved that the voting machines were, in fact, simple computers by forcing them to play chess [10], contradicting the manufacturers claim that they were machines done just for voting: *"this particular TV broadcast (...) was absolutely decisive"* (Int. 2). Once the problem reached the public audience, the government's response was notably inadequate, caught off guard by the magnitude of the situation, *"the government couldn't explain why they had used voting machines and how they could verify that they gave the rights results"* (Int. 3). This lack of preparedness was compounded by the lack of technical competences of those tasked with providing explanations, *"you have a ministry filled with people that have studied law"* (Int. 5). Additionally, the reluctance of the vendor to open up the system for scrutiny in order to protect their business interests further hampered efforts to address the concerns raised by the hacktivists. This left the hacktivists without credible counterparts to engage with regarding their legitimate claims. *"If you cannot give a robust answer to that kind of things, then trust will erode"* (Int. 6).

In response, the government had to act and created two external commissions to scrutinize the electoral process, "When something goes wrong, in Dutch politics, they usually appoint a committee, wise men and women, and they study the situation, and they give an advice" (Int. 3). The first commission, known as the "Voting Machines Decision-Making," was tasked with examining the rationale behind the adoption of voting machines. Simultaneously, the second commission, "Election Process Advisory," focused on charting the course for future electoral procedures in the Netherlands [49]. Ultimately, the outcome of this comprehensive review was the decision to discontinue the use of voting machines in 2008, opting instead to revert to the traditional method of paper ballots. This decision remains unchanged to date.

### 4.6 The aftermath

Interestingly, the legal decision to withdraw the voting machines was not primarily driven by concerns over their technical vulnerabilities of voting machines or the potential for fraud. Instead, it centered on the breach of legal requirements highlighted by one of the identified weaknesses. *"Secrecy of the vote is constitutionally protected, whereas the integrity of (...) the vote is not. In order for* (hacktivists) *to have a legal remedy, (...) they had to prove that the machines were violating the Constitution"* (Int. 8). The main argument for the decision was that the secrecy of the vote could be compromised due to a distinct noise produced by the voting machines when a ballot containing a special character (e.g., ë) was cast. This unique sound could be registered using specialized audio equipment at a distance of up to 25 meters [31]. *"If they wouldn't come up with this idea that you can read the waves depending on who you are voting (...), they wouldn't have been successful in court"* (Int. 8). This legal loophole underscored the significance of constitutional protections surrounding the secrecy of the voting process, ultimately leading to the decision to abandon the use of the machines.

The impression of those involved in the crisis is that the decision to withdraw the use of voting machines was adequate, *"it was irresponsible to use this technology"* (Int. 3*), "looking back at it from the point of view of the things that we've learned since then. cyber security wasn't a big issue at that point…"* (Int. 1), *"We couldn't give a robust response, because* (hacktivist) *was completely right"* (Int. 6). Looking back, the assumption is that things could have been done in a very different manner: *"The problem that aroused when* (hacktivists) *looked at the machines was that there was no security, there were no security requirements, as such. Because nobody really had thought that such requirements were needed. Because the thought was, we trust the vendors, they make the machines, they program (...). We have requirements regarding the municipalities that the machines can only be used for voting. And that during the periods that are no elections, they have to be guarded by the municipalities and nothing can be done with them. And we trust the municipalities. So, if you trust everything, you don't get into the mindset that you probably have to think about things you cannot trust, or that there are risks, or threads that you have to research and find out if the trust that you had must be thought through again because things can change"* (Int. 6).

However, this perspective seems limited to the expert community since the debate regarding the trustworthiness of voting machines did not reach a wider audience. Despite the media and Parliament debates, data from the Dutch Parliamentary Election Studies of 2006 and 2010 does not unequivocally suggest a severe erosion of trust in voting machines among Dutch voters. An analysis of these surveys [44] reveals a nuanced picture: while the percentage of voters expressing a high level of trust in voting machines declined from 84% to 69%, a significant proportion of citizens still prefer voting machines over paper ballots. In 2010, 47% of respondents favored voting machines, compared to 24% who preferred paper voting. Moreover, when directly compared, 27% of voters deemed voting machines more trustworthy than paper ballots, while 22% expressed greater trust in paper ballots than voting machines. *"It wasn't a big issue, in general, in the general public. People, in general, trusted the voting machines, even after"* (Int. 1).

Similarly, the decision to withdraw the voting machines was met with discontent among municipalities, reflecting concerns about increased workload *("They were angry because they said 'it's a lot more work for us',"* Int. 8). However, this dissatisfaction seemed to stem from a perception that the threat posed by the voting machines was largely hypothetical and not grounded in reality, *"they thought that there wasn't really a threat. It was a fairy theory (...) it still is regarded by a lot of the municipality as something that in theory could take place, but it's not a real threat*" (Int. 6), and they are still pushing for their return *"municipalities still want voting machines, because to them, elections are a very messy process"* (Int. 5).

## 5 Discussion

The significance of the case at hand extends across multiple dimensions. Firstly, within the field of electoral technologies, the Dutch case stands as a prominent reference point for researchers [45] [53], offering valuable insights into the management of electronic voting systems. The experience of the Netherlands created an important precedent in the community, for example, highlighting the need to avoid black-box voting systems and promote transparency to build trust and help manage distrust. However, its impact goes beyond academia; for instance, the German constitutional court's ruling against voting machines and stressing the need for transparency in the electoral process without specialist technical knowledge[1].

This research focuses on examining the failure of voting machines in the Netherlands through the lens of trust and distrust formation and management. Trust and distrust are strategies to deal with the complexity lying between rationality and emotions [70] that are pumped with facts and experience. As Kühne [39] suggests, trust hinges on the absence of negative inputs to discuss the trustworthiness of the object of trust, while distrust can emerge with a good concatenation of negative inputs. These concepts align seamlessly with the narrative previously outlined. The perceived convenience of the technology in use for various levels of administration, coupled with a general lack of technical expertise and a prevailing culture of trust in electoral authorities and institutions, veiled the potential risks associated with using voting machines. Yet, it took merely a small but well-prepared group of hacktivists to disrupt the system entirely.

When revisiting the case, it becomes evident that the biggest hack of the hacktivist group was not towards the machines that were easy to hack [26] but to the overall electoral management system. Their actions revealed a series of inadequacies in the process of adoption and management of voting machines. The rise of distrust stemmed not only from the machines' demonstrated lack of reliability but predominantly from the bad security management, the lack of legal provisions in cybersecurity, or how voting machines were physically safeguarded.

Based on the previous information, a set of factors contributing to trust or distrust can be identified and distributed using the proposed theoretical differentiation between trust and distrust, allowing to detect elements that simultaneously contribute to the creation of trust and distrust, as well as others having negligible impact on either. In Table 1, the main elements in play are shortlisted, describing their impact on the creation of trust and/or distrust.

**Table 1.** Factors contributing to trust and/or distrust.

| Factor (in alphabetical order) | Contributes to Trusting/ Distrusting | Reason |
| --- | --- | --- |
| Expert debate | None | The complexity of the debate kept it far from most of the population, who did not change their position regarding trusting the democratic system or the voting machines. The lack of political controversy on the topic helped keep it largely unnoticed by a broad audience. |
| High level of trust in institutions | Trust | The traditionally high level of trust in institutions among the Dutch citizenry facilitated the widespread acceptance and unquestioned adoption of voting machines but also mitigated the potential negative impact of the case in future elections. |
| Lack of critical approach towards Voting Machine | Both | The absence of a critical generalized attitude regarding voting machines fostered a false sense of trust, wherein trust was posited in something not deserving it, consequently seeding the roots of distrust. |
| Legal amendments after the crisis | Trust | Reforming legislation and ensuring transparency and openness in elections allowed the integrity of future electoral management and democracy. |
| Long-term reaction of the administration | Trust | The resolution of the problem, with the withdrawal of voting machines allowed protecting the trustworthiness of the electoral management and democracy. |
| Media and Hacktivist campaign | Distrust | The hacktivists' communication strategy and the collaboration of media resulted in an effective domination of the discourse, fueling the creation of distrust in the voting machines. |
| Outdated and inadequate legal framework | Both | The legal framework was not prepared to address the emerging problem as the security concept in use did not align with the actual threats. While the legal framework contributed to legitimizing the system in the eyes of the public, it ultimately became a source of distrust when its inadequacy in safeguarding against threats was exposed. |
| Positive previous experience | Trust | The successful experience using voting machines prevented critical discourses from appearing. |
| Previous activities of the hacktivists | Both | The hacktivists' prior activities gave them recognition and legitimacy amongst certain key actors. While they were bringing distrust to the voting machines, it became evident their |

---

[1] See: https://www.ndi.org/e-voting-guide/examples/constitutionality-of-electronic-voting-germany

|  |  | contribution to increasing the trustworthiness of the overall democracy. |
|---|---|---|
| Short-term reaction of the administration | Distrust | The administration's lack of capacity to effectively address the concerns raised by the hacktivists generated distrust in the system and cast a general doubt on the suitability of using voting machines. It became evident that the administration was not ready to confront the problem. |
| Weaknesses of the voting machine | Distrust | The proven weaknesses of the voting machines were a legitimating factor for the hacktivist claims. |
| Withdrawal of the system | None | The withdrawal of voting machines did not disrupt the democratic system itself, indicating that issues of trust or distrust in democracy might be rooted in other underlying sources. |

The preceding factors can be visually represented in an XY Axis grid (Figure 1), where trust and distrust are displayed. It is important to clarify that the positions on the grid do not mean intensity but rather their alignment within the quadrant.

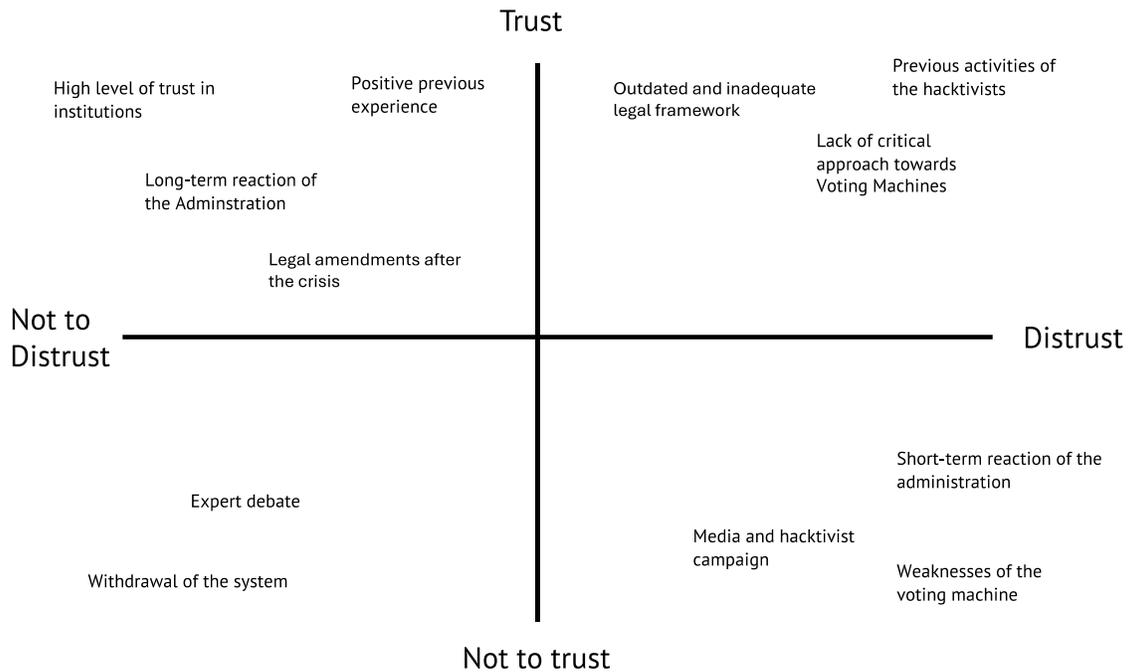

Figure 1. Distribution of factors in Trust and Distrust axis

The Dutch case leaves some interesting takeaways on how trust and distrust intertwine in relation to the management of electoral technologies. First of all, the trustworthiness of technology is at the center of the problem but is inserted in a socio-technical environment, in a permanent dialogue with the cultural and administrative context in which it operates. Hacktivists' legitimate distrust regarding voting machines might not have been successful without the perfect storm that was unchained thanks to the administrative management of the situation and a good dissemination strategy supported by the media.

Secondly, the case shows how trust and distrust can occur in parallel. The legal framework surrounding the Dutch voting machines exemplifies this duality: it initially helped trust in the reliability of the machines, yet when it failed to address the concerns raised by hacktivists, it became a source of distrust. Democracy heavily relies upon the existence of societal agreements that are secured by the performance of public administrations. However, the speed of technological transformation often outruns the administrative capacities, creating legal and administrative gaps where unexpected events can occur and that, as shown in this case, can shake the foundations of trust.

Thirdly, the dynamics of trust and distrust creation have different impacts across societal levels. While the discussions surrounding voting machines deeply affected the ministry's core, its impact on average citizens was limited. Surveys conducted after the withdrawal of voting machines [47] indicate that voters continued trusting in voting machines even after proven untrustworthy, perhaps due to a transference of trust in institutions and a lack of technical expertise to fully comprehend the issue. This sets a dangerous precedent, *people will use insecure systems if they feel or think they are secure* [55], highlighting the relevance of social factors for the production of trust. Moreover, the Dutch administration's ultimate decision reflects this understanding: the voting system must be transparent and understandable to every citizen, without requiring specialized knowledge, to prevent them from placing trust in something inherently untrustworthy. This reinforces the administration's central role in establishing secure technological environments and the need for experts to engage in decision-making processes involving critical sectors. Additionally, it contributes to explaining why electoral technologies are not widely adopted, aligning with the middleman paradox proposed by Mahrer and Krimmer [49].

Finally, convenience is crucial in creating trust in electoral technologies. A primary argument favoring technology adoption is its ability to simplify activities or processes. However, every technological adoption entails a series of trade-offs that must be carefully evaluated before implementation. These trade-offs typically manifest as humanly-created risks that we accept in exchange for the benefits offered by the technology [7]. In this case, the risk was the potential for vendors to manipulate election results (albeit unlikely, according to interviewees' perspectives), with no means of definitively proving such tampering. Therefore, convenience can be understood as a gateway to technology acceptance on the condition that overall experiences remain positive and risks are perceived as manageable.

# 6  Conclusions

The Dutch case offers valuable insights into the intricate interplay between trust and distrust in the management of electoral technologies: 1) it highlights the centrality of technology's trustworthiness and the necessary role of the socio-technical context in which it is inserted; 2) it demonstrates how trust and distrust can coexist simultaneously; 3) it shows how trust and distrust dynamics have different impacts across societal levels and 4) it emphasizes the role of public administration in establishing secure technological environments as a foundation of long-lasting trust.

Beyond these insights and the specific takeovers that the electoral community has already taken from previous approaches to this case [45] [54], this research offers a broader reflection on the interaction between trust and distrust in data-driven technologies [21]. While trust is something "to be constructed" based on elements such as convenience, habit, previous experience, and updated legal regulations, distrust is something "to be managed," and that can be raised with an active engagement from interested actors (even if it can be for legitimate reasons as in the case presented). Following the previous, public administrations should consider developing different strategies for 1) building trust and 2) dealing with distrust.

This nuanced relation between trust and distrust also reveals an interesting paradox: trust can be a source of distrust under certain conditions, and vice versa. For example, the hacktivist group aimed to build trust in the democratic system by creating distrust in the voting machines. Similarly, the legal framework was initially helping to build trust in the voting machines but later became a source of distrust. Other important aspects to consider include the role of convenience in fostering technology adoption and hindering the emergence of distrust discourses, as well as the transference of trust (and distrust) between different actors. These areas offer avenues for future research, such as investigating the role of convenience in trust formation across various technologies and exploring how technological failures may rely on trust transference processes in scenarios of institutional distrust.

## ACKNOWLEDGMENTS
Data collection for this paper was funded by the Polish National Research Center's grant Miniatura 3 - 2019/03/X/HS6/01688. The data analysis and development of the publication has been funded by the Polish National Research Center's grant OPUS-20 - 2020/39/B/HS5/01661 and EU H2020 MSCA Program, grant agreement no. 101038055. The author would like to extend its gratitude to the interviewees who shared their insights on this case, to the peer-reviewers who read and commented on my work, and to Robert Krimmer and Leontine Loeber for their valuable comments on previous iterations of this work. AI was used in this text for reviewing grammar.